\documentstyle[sprocl,epsfig]{article}

\def\be{\begin{equation}}
\def\bea{\begin{eqnarray}}
\def\eea{\end{eqnarray}}
\newcommand{\mH}{\mbox{$m_{\mathrm{H}}$}}
\newcommand{\sqrts}{\mbox{$\sqrt {s}$}}
\newcommand{\Gcs}{\mbox{${\rm GeV/}c^2$}}

\newcommand{\mh}{\mbox{$m_{\mathrm{h}}$}}
\newcommand{\mA}{\mbox{$m_{\mathrm{A}}$}}
\newcommand{\ee}{\mbox{$\mathrm{e}^{+}\mathrm{e}^{-}$}}
\newcommand{\ra}{\mbox{$\rightarrow$}}
\newcommand{\sba}{\mbox{$\sin ^2 (\beta -\alpha)$}}
\newcommand{\tanb}{\mbox{$\tan \beta$}}
\newcommand{\cba}{\mbox{$\cos ^2 (\beta -\alpha)$}}
\newcommand{\tautau}{\mbox{$\tau^{+}\tau^{-}$}}
\newcommand{\bb}{\mbox{$\mathrm{b} \bar{\mathrm{b}}$}}
\newcommand{\mHpm}{\mbox{$m_{\mathrm{H}^{\pm}}$}}
\newcommand{\Hp}{\mbox{$\mathrm{H}^{+}$}}
\newcommand{\Hm}{\mbox{$\mathrm{H}^{-}$}}
\newcommand{\csbar}{\mbox{$\mathrm{c} \bar{\mathrm{s}}$}}
\newcommand{\cbars}{\mbox{$\bar{\mathrm{c}}\mathrm{s}$}}
\newcommand{\tp}{\mbox{$\tau^+$}}
\newcommand{\tm}{\mbox{$\tau^-$}}
\newcommand{\nubar}{\mbox{$\bar{\nu}$}}

\begin{document}
\begin{titlepage}
\def\thefootnote{\fnsymbol{footnote}}       

\begin{center}
\mbox{ } 

\vspace*{-4cm}


\end{center}
\begin{flushright}
\Large
\mbox{\hspace{10.2cm} hep-ph/0004015} \\
\mbox{\hspace{10.2cm} IEKP-KA/2000-06} \\
\mbox{\hspace{12.0cm} April 2000}
\end{flushright}
\begin{center}
\vskip 1.0cm
{\Huge\bf
Higgs Boson Searches

at LEP
\smallskip

Up To \boldmath$\sqrts=202$\unboldmath\ GeV}
\vskip 1cm
{\LARGE\bf Andr\'e Sopczak}\\
\smallskip
\large University of Karlsruhe

\vskip 2.5cm
\centerline{\Large \bf Abstract}
\end{center}

\vskip 1cm
\hspace*{-3cm}
\begin{picture}(0.001,0.001)(0,0)
\put(,0){
\begin{minipage}{16cm}
\Large
\renewcommand{\baselinestretch} {1.2}
The latest preliminary combined results of the
Higgs boson searches from the LEP experiments ALEPH, DELPHI,
L3 and OPAL are presented. A general scan of the MSSM parameters is performed
and leads to stringent lower mass limits and for the first time excludes low 
$\tan\beta$ values.
\renewcommand{\baselinestretch} {1.}

\normalsize
\vspace{4cm}
\begin{center}
{\sl \large
Presented at the Seventh International Symposium on Particles, Strings, and
Cosmology, PASCOS--99, Granlibakken, USA, Dec. 10--16, 1999
\vspace{-6cm}
}
\end{center}
\end{minipage}
}
\end{picture}
\vfill

\end{titlepage}


\newpage
\thispagestyle{empty}
\mbox{ }
\newpage
\setcounter{page}{1}

\title{Higgs Boson Searches at LEP Up To 
\boldmath$\sqrts=202$\unboldmath\ GeV}

\author{Andr\'e Sopczak}

\address{Karlsruhe University \\E-mail: andre.sopczak@cern.ch} 


\maketitle\abstracts{The latest preliminary combined results of the
Higgs boson searches from the LEP experiments ALEPH, DELPHI,
L3 and OPAL are presented. A general scan of the MSSM parameters is performed
and leads to stringent lower mass limits and for the first time excludes low 
$\tan\beta$ values.}

\section{Standard Model Higgs Boson}
\begin{figure}[htb]
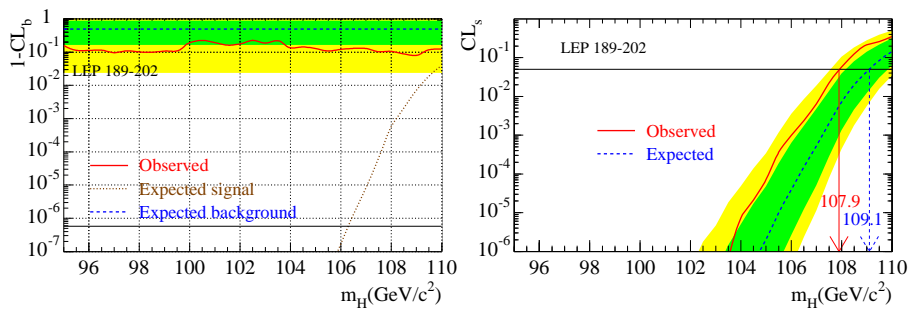

\begin{center}
\epsfig{figure=lep202_mclb.epsi,width=0.48\textwidth}
\epsfig{figure=lep202_cls.epsi,width=0.48\textwidth}
\vspace*{-1mm}
\caption[]{\small 
Confidence levels $CL_b$ and $CL_s$ from combining the data collected 
by the four LEP experiments at energies from 189 to 202~GeV.
The solid curve is the observed result and 
the dashed curve the expected median.
The shaded areas represent the symmetric $1\sigma$ and $2\sigma$ 
probability bands. 
The horizontal line at $1-CL_b=5.7\times 10^{-7}$ indicates the level 
for a $5\sigma$ discovery, which shows that a 106~\Gcs\ Higgs boson
could have been discovered,
and the intersections of the curves with 
the horizontal line at $CL_s=0.05$ give the mass limits at 95\% CL.
\label{sm-cl}}
\end{center}
\vspace*{-2mm}
\end{figure}

\begin{figure}[htb]
\begin{center}
\epsfig{figure=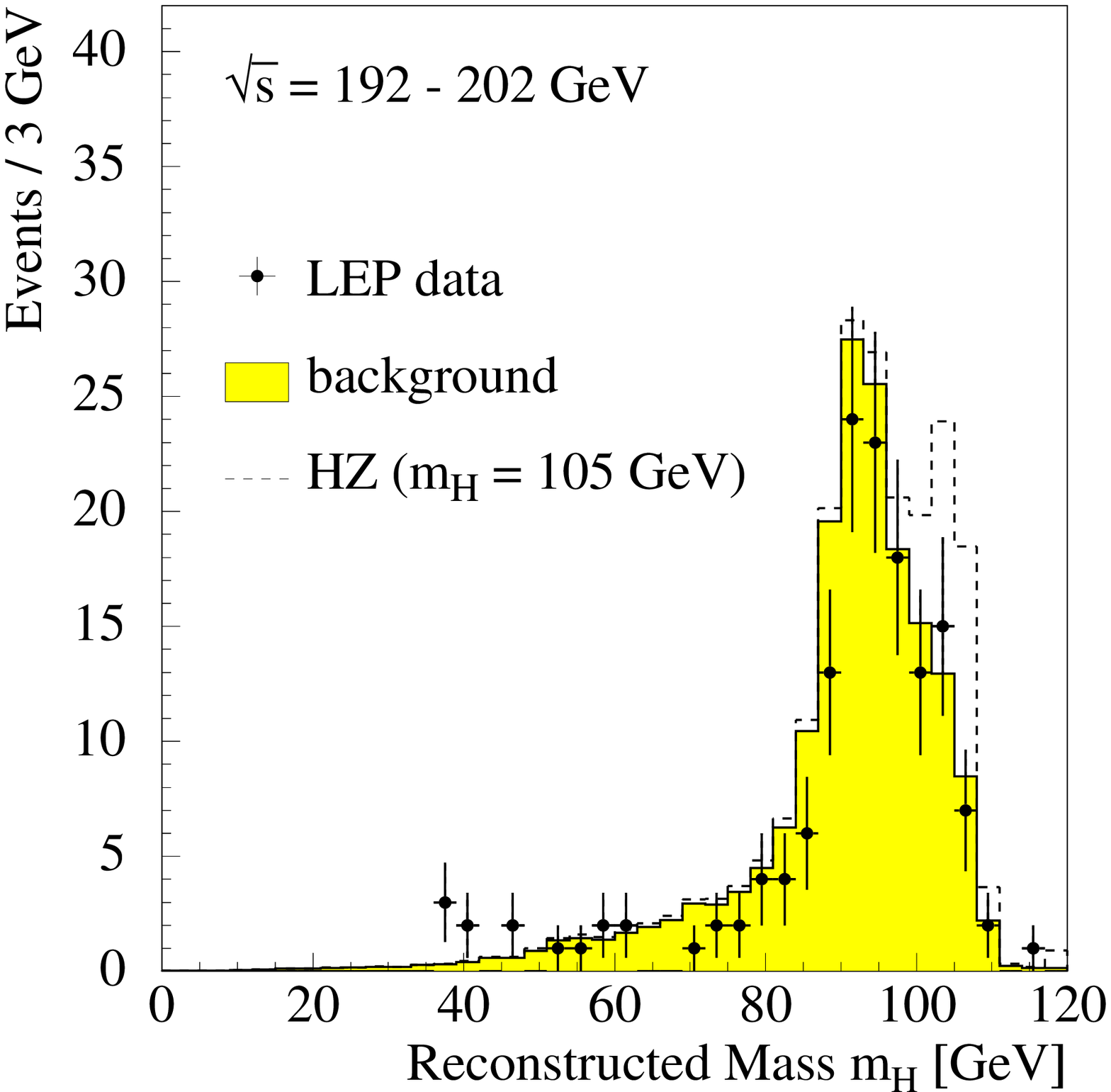,width=0.48\textwidth}
\epsfig{figure=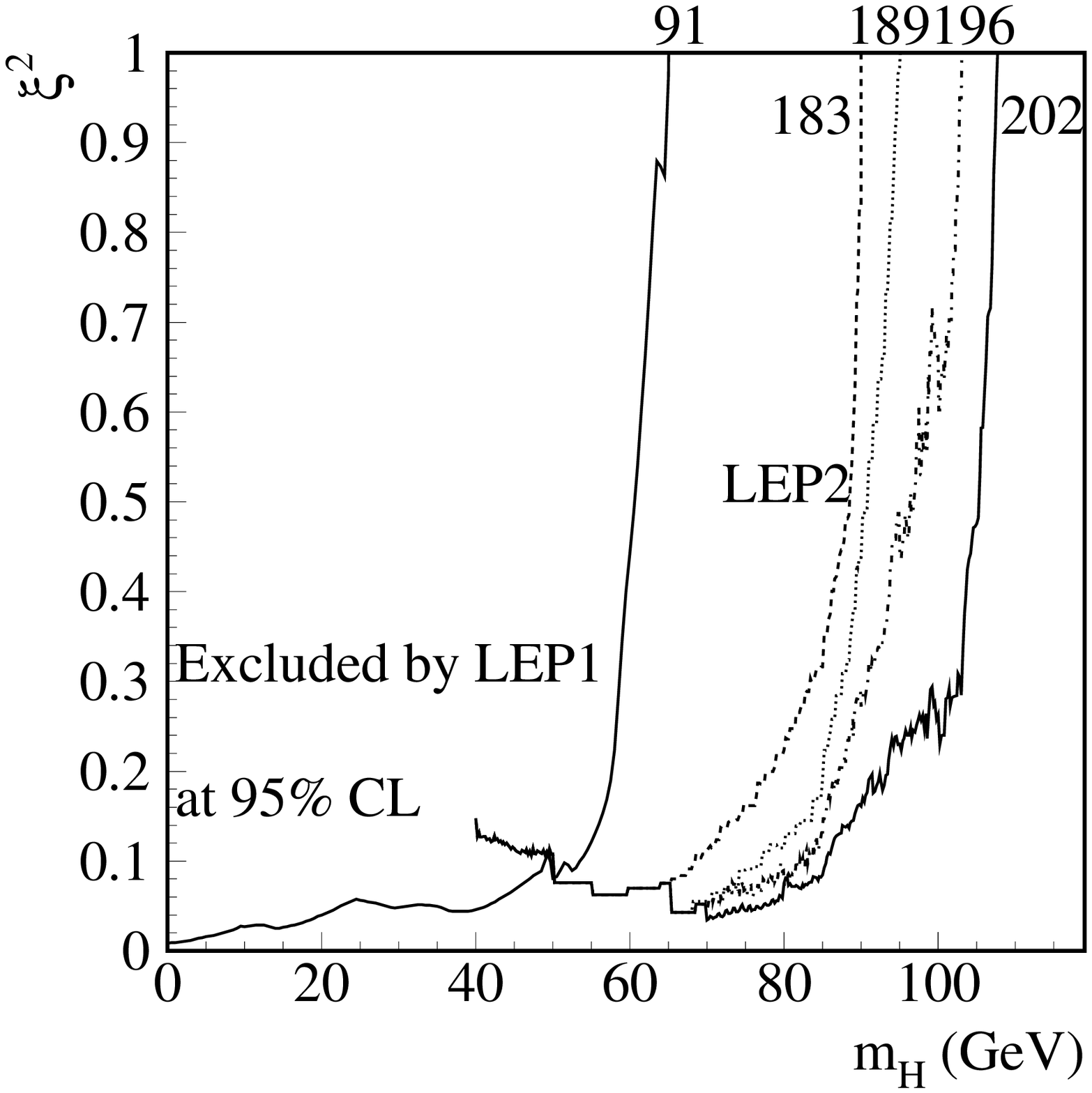,width=0.48\textwidth} 
\caption[]{\small 
Left: Distribution of the reconstructed SM Higgs boson mass in searches 
conducted at energies between 192 and 202 GeV.
The figure displays the data
(dots with error bars), the predicted SM background (shaded histogram)
and the prediction for a Higgs boson of 105~\Gcs\ mass
(dashed histogram). 
The number of data events selected for this figure is 148, while 175 
are expected from SM background processes. 
A signal at 105~\Gcs\ mass would contribute with 36 events. 
Right:
The 95\% CL upper bound on $\xi^2$ as a function 
of \mH, where $\xi= g_{\rm HZZ}/g_{\rm HZZ}^{\rm SM}$ is the HZZ coupling 
relative to the SM coupling, including combined LEP1~\cite{lep1} 
and LEP2~\cite{moriond00} results. The excluded regions are shown for
data including up to 91, 183, 189, 196, and 202 GeV center-of-mass energies.
\label{sm-mass}}
\end{center}
\end{figure}

The search for Higgs bosons at LEP shows no indication of a signal.
This review includes the data from 
the outstanding performance of the LEP accelerator
in 1999 with a collected luminosity of
about 900~pb$^{-1}$ at $\sqrts=192$ to 202~GeV~\cite{moriond00}.
The confidence levels $CL_b$ for a signal observation 
and $CL_s$ for setting mass limits are shown in Fig.~\ref{sm-cl}.
The resulting Standard Model (SM) Higgs boson mass limit is
107.9~\Gcs\ at 95\% CL.
The reconstructed mass distribution is shown 
in Fig.~\ref{sm-mass}. In extensions of the SM 
the HZZ coupling might be weaker and thus the production cross section 
is reduced. Figure~\ref{sm-mass} shows limits on the reduction factor
at 95\% CL. Even if the SM cross section is reduced by a factor three,
a Higgs boson mass up to 103~\Gcs\ is excluded.

\section{MSSM Benchmark Results}

\begin{figure}[htb]
\begin{center}
\begin{minipage}{0.48\textwidth}
\epsfig{figure=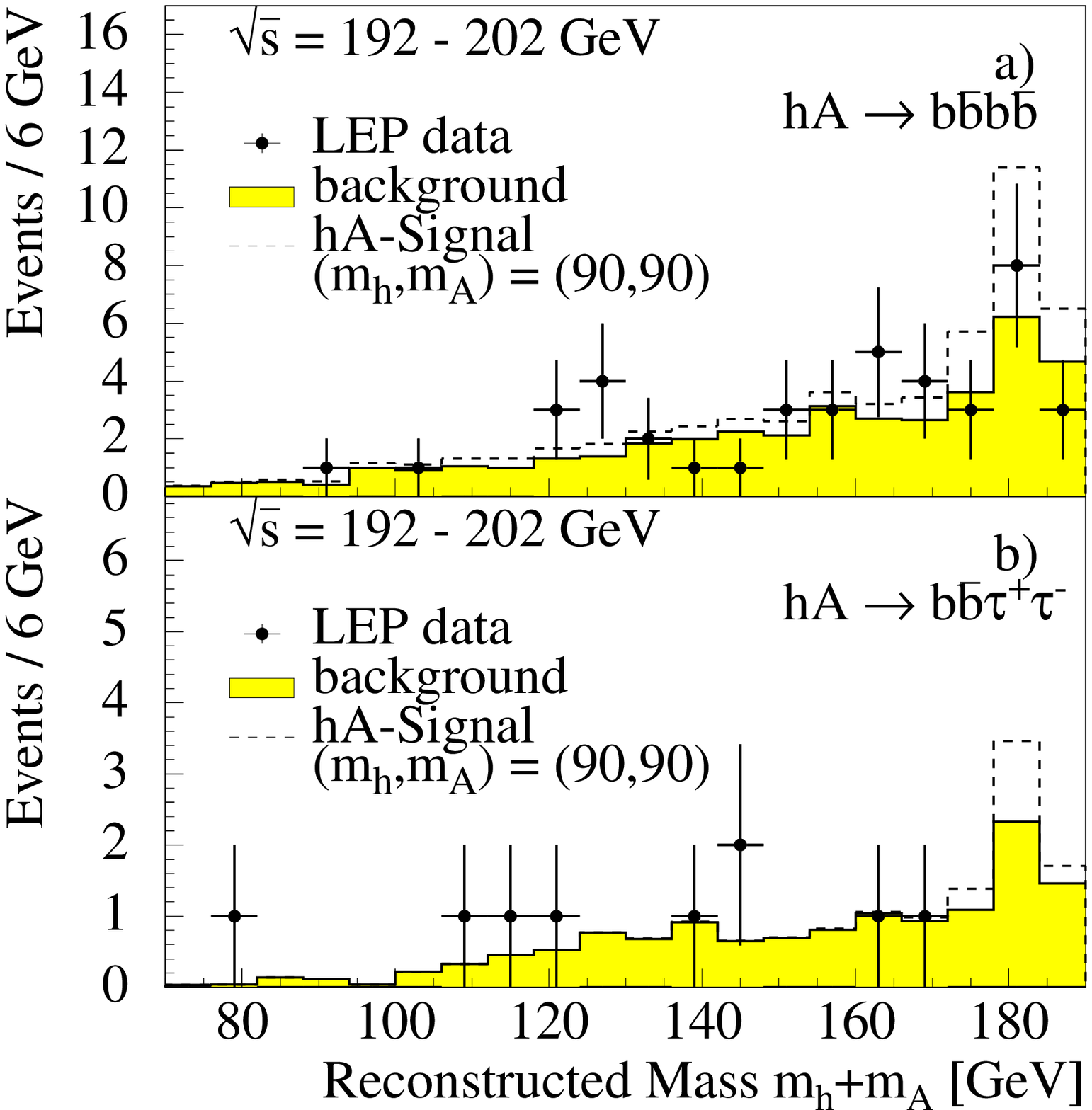,width=1.0\textwidth}
\end{minipage}
\begin{minipage}{0.48\textwidth}
\epsfig{figure=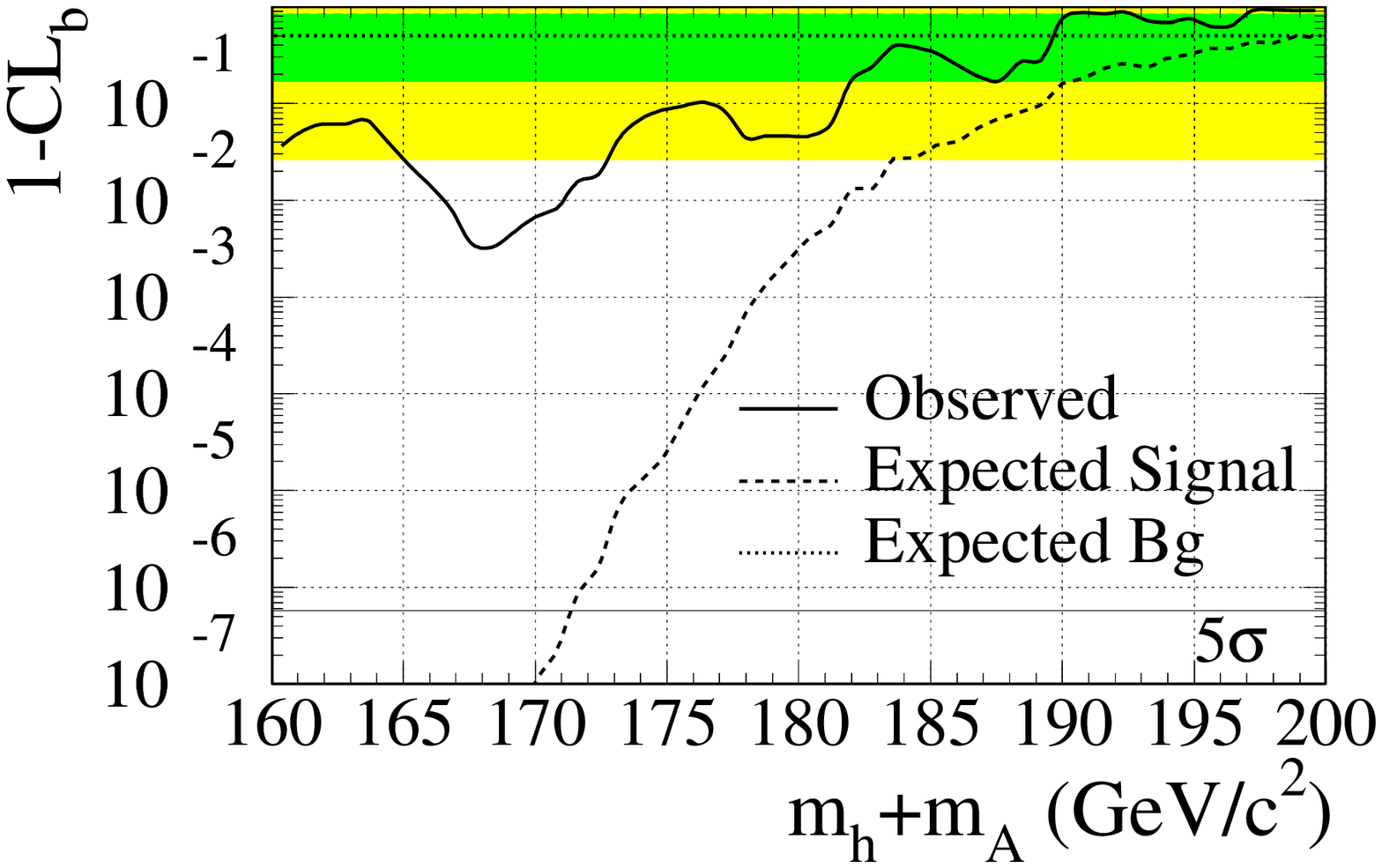,width=1.1\textwidth}

\vspace*{-1cm}
\epsfig{figure=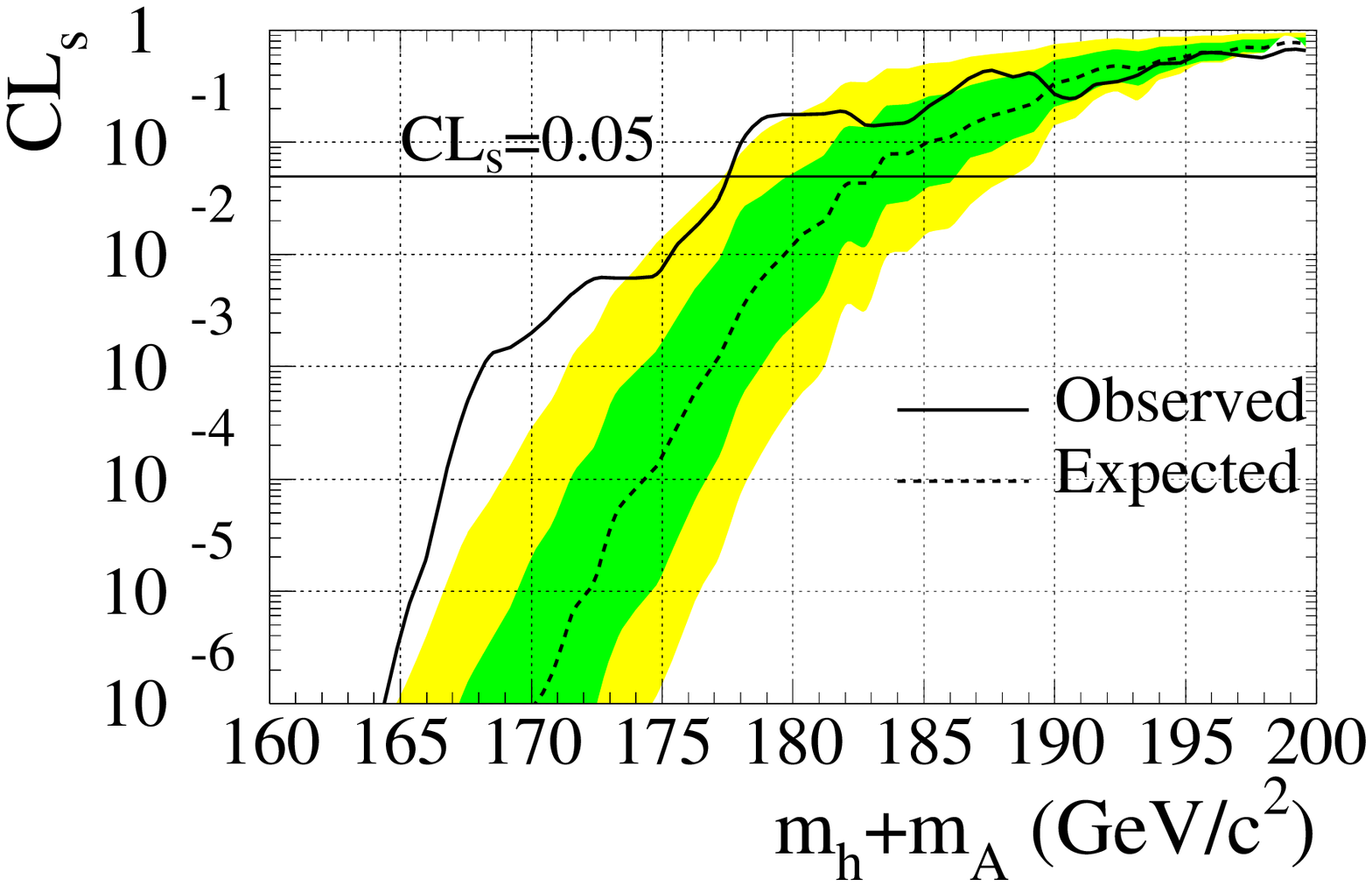,width=1.1\textwidth}
\end{minipage}
\caption[]{\small 
Left:
Distribution of the reconstructed Higgs mass sum $\mh+\mA$ 
in searches
for the MSSM process \ee\ra~hA~\ra~\bb\bb\ and \bb\tautau. 
The data of the four LEP experiments
collected at energies from 192 to 202~GeV are added.  
The figure displays the data
(dots with error bars), the predicted SM background (shaded histogram)
and the prediction for a MSSM signal with $\mh\approx\mA=90$~\Gcs\
(dashed histogram) for which the maximal cross-section ($\cba=1$) has been
assumed. In the hA~\ra~\bb\bb\ case the mass sum $\mh+\mA$ is shown only
for the dijet pair with the smallest mass difference $\mid$\mh-\mA$\mid$.
The number of data events entering the upper (lower) 
figure is 42 (9) for 39.5 (11.2) SM background events expected. 
A signal with the above characteristics would produce 14.7 (0.8) events.
Right:
The confidence level $CL_b$ and $CL_s$ 
as a function of $\mh+\mA$, for the \mh-max benchmark 
and the particular case $\mh\approx\mA$
(where only the \ee\ra~hA process contributes since $\sba\approx0$). 
The straight dotted line at 50\% and the shaded bands represent the 
median result and the symmetric $1\sigma$ 
and $2\sigma$ probability bands.
The solid curve is the observed result and the dashed curve shows the 
expected median.
The horizontal line at $1-CL_b=5.7\times 10^{-7}$ indicates the level 
for a $5\sigma$ discovery and the intersections of the curves with the 
horizontal line at $CL_s=0.05$ give the limit on $\mh+\mA$ 
at 95\% CL.

\label{mssm-mass}}
\end{center}
\end{figure}

\begin{figure}[htp]
\begin{center}
\epsfig{figure=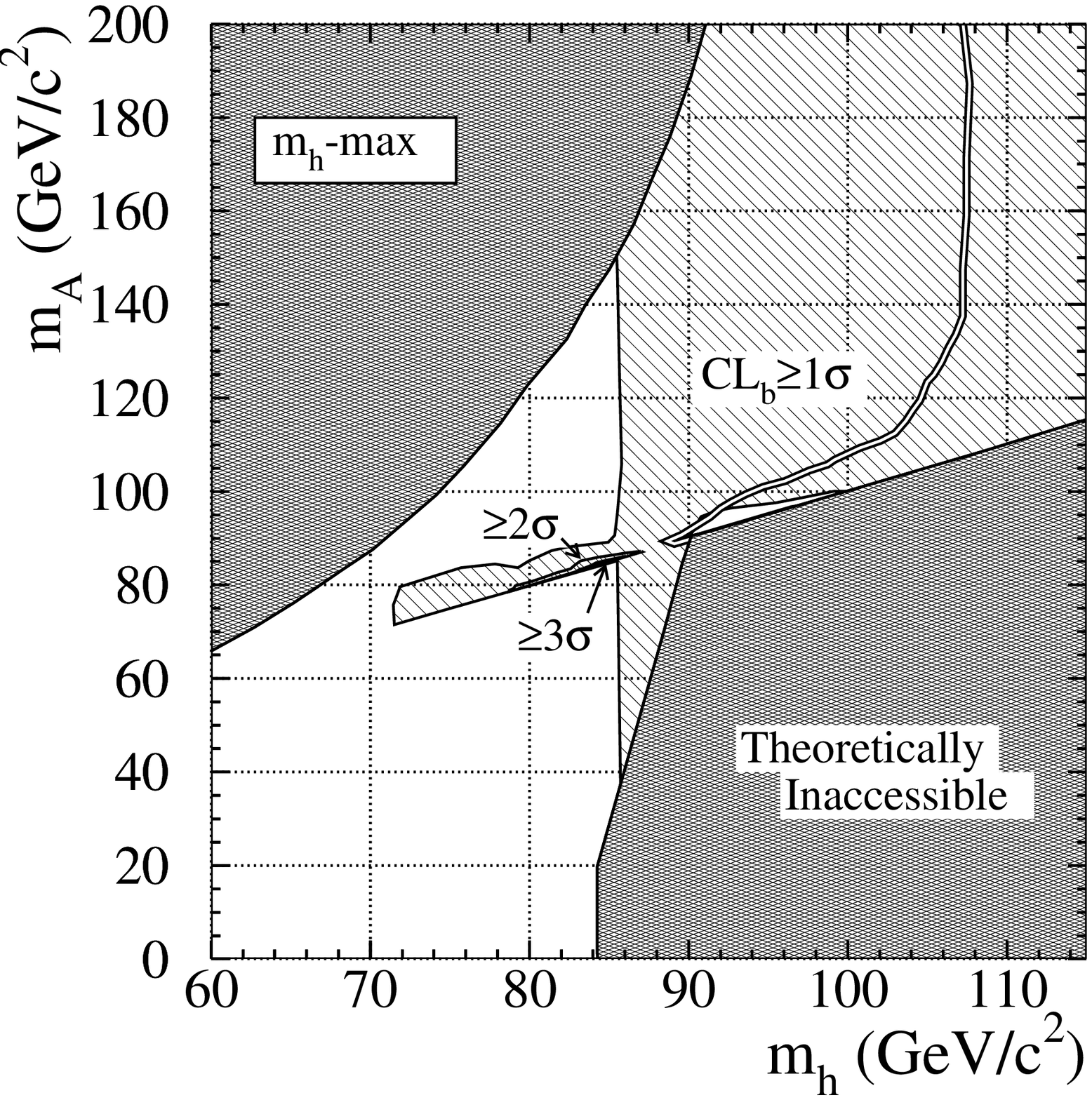,width=0.48\textwidth}
\epsfig{figure=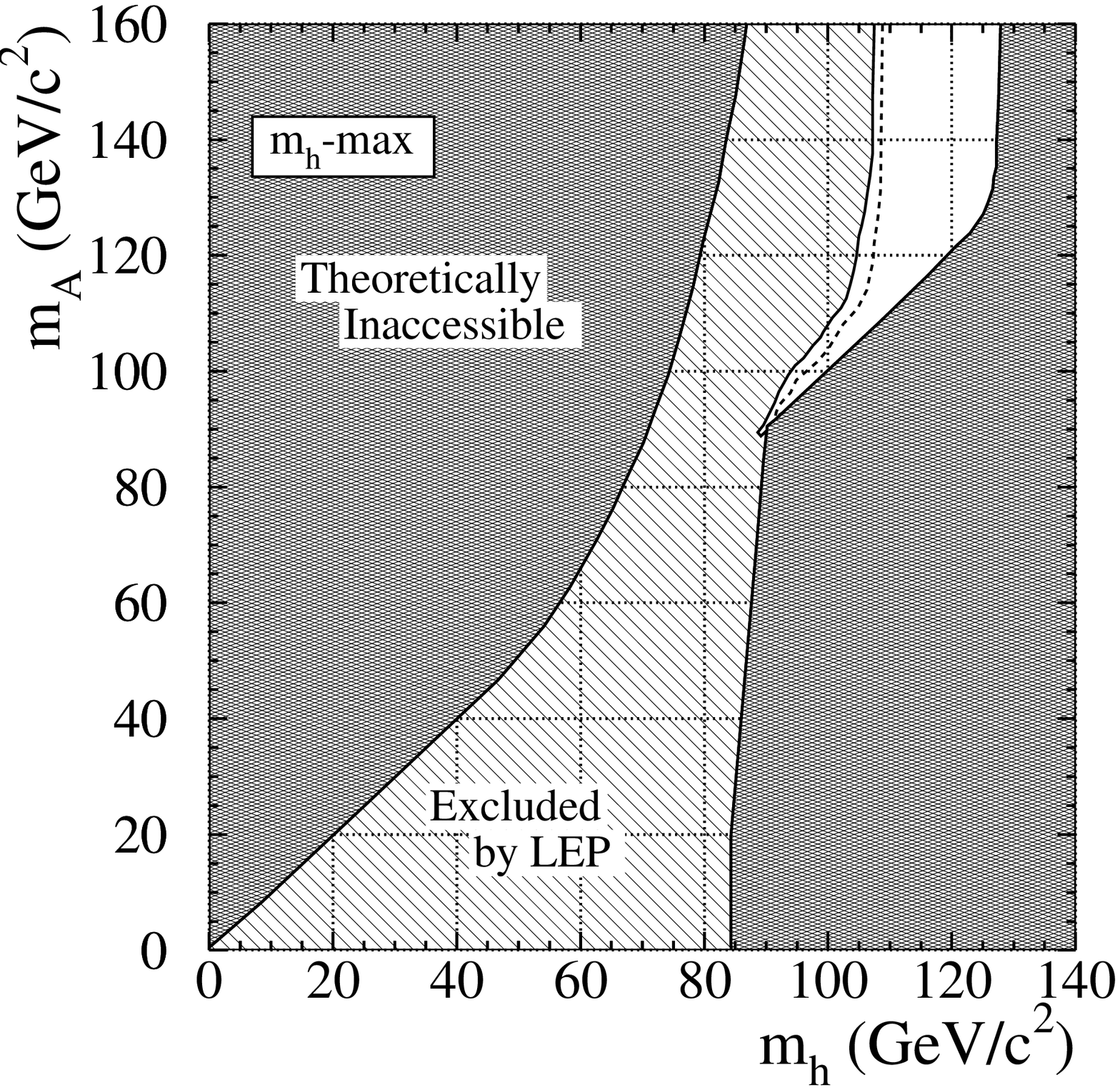,width=0.48\textwidth} 
\epsfig{figure=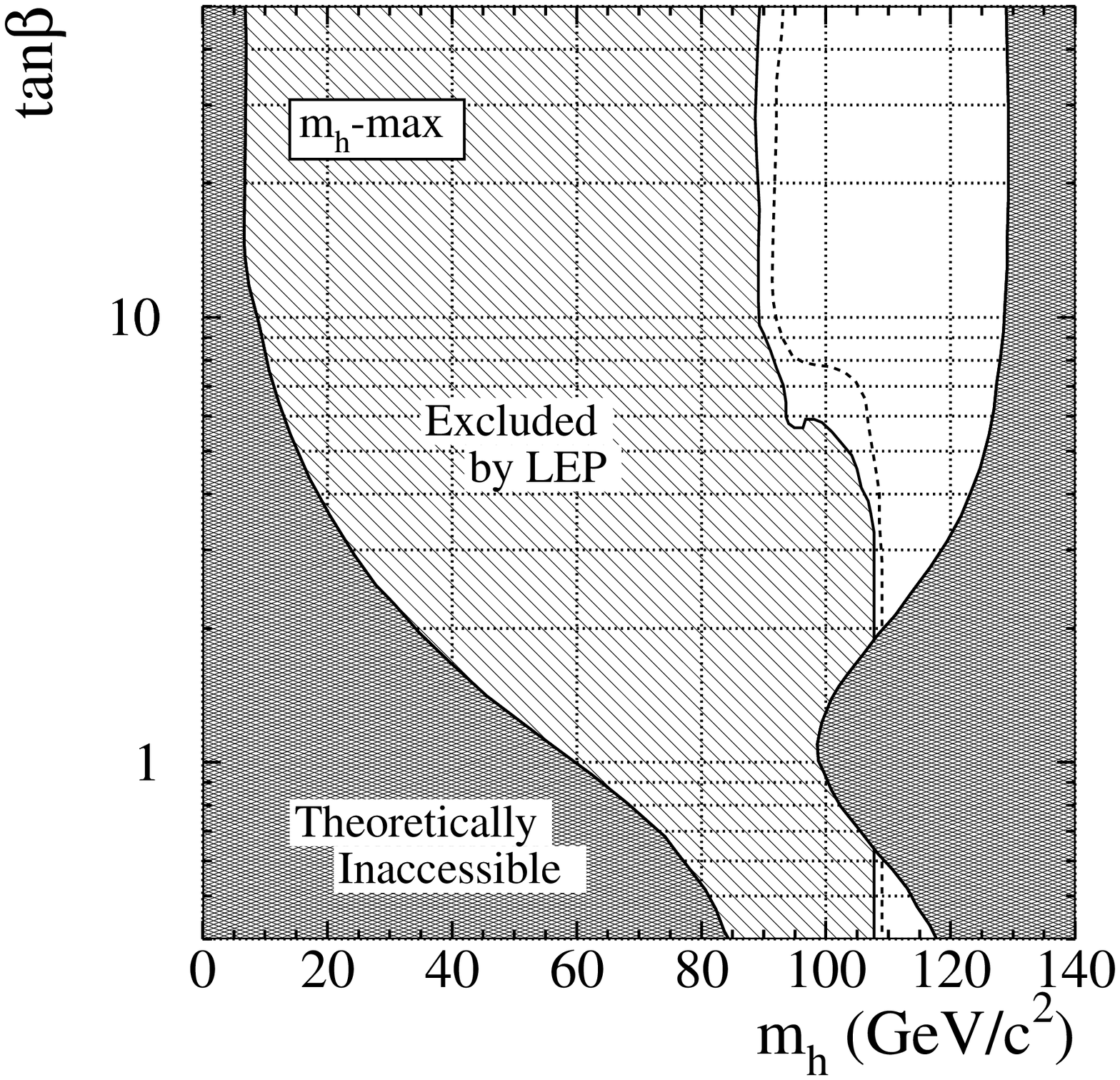,width=0.48\textwidth} 
\epsfig{figure=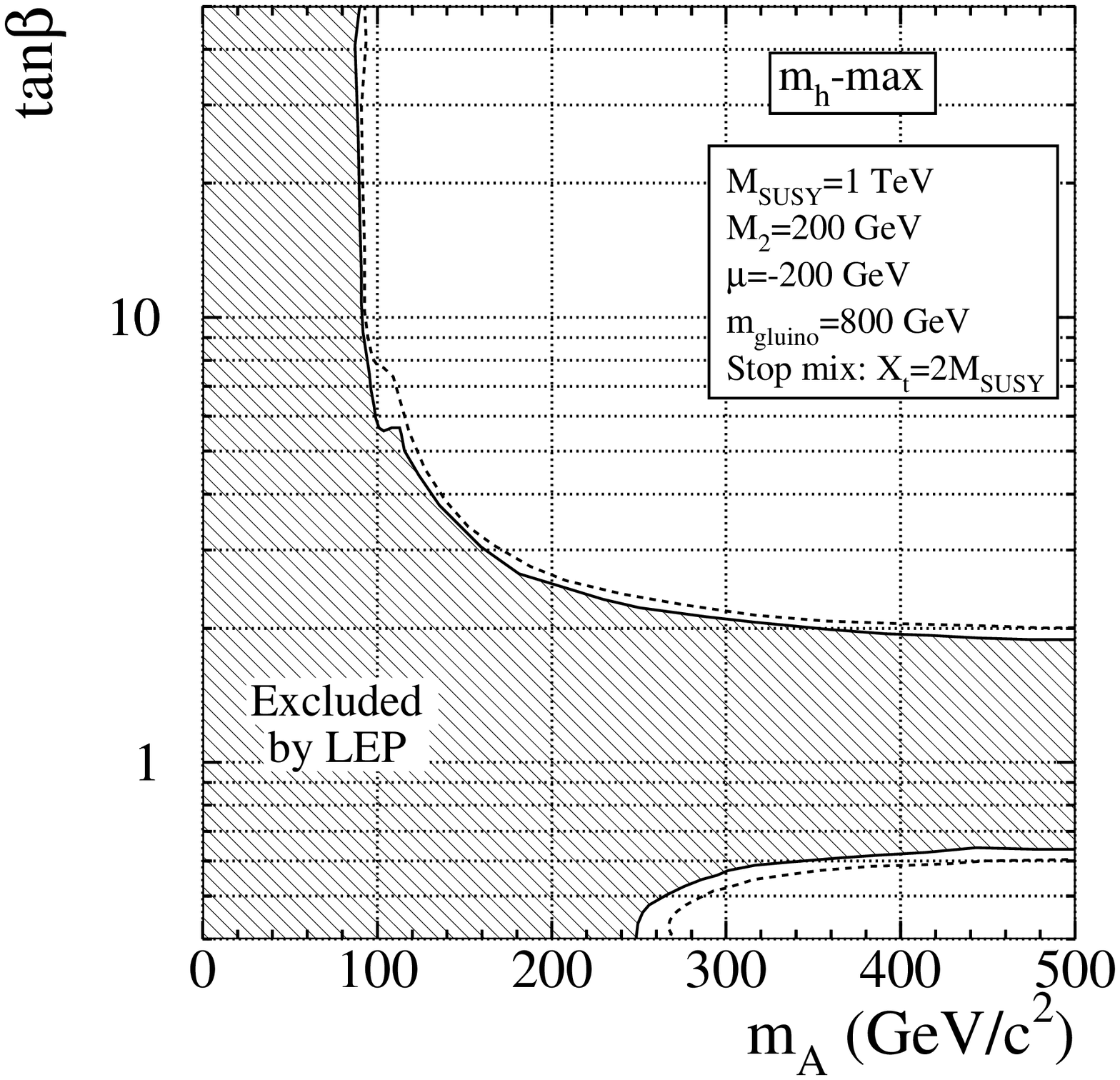,width=0.48\textwidth} 
\caption[]{\small 
Upper left:
Distribution of the discovery confidence level $1-CL_b$ for the \mh-max 
benchmark, projected onto the ($\mh,\mA$) plane from combining the data of 
the four LEP experiments at energies from 192 to 202~GeV. 
In the white domain the observation either shows a deficit or is 
less than $1\sigma$ above the background prediction;
in the domains labelled $1\sigma$, $2\sigma$ and $3\sigma$ the observation 
is between 1 and 2$\sigma$, 2 and 3$\sigma$ and larger than 3$\sigma$, 
respectively, of the prediction.
The other plots show the 95\% CL bounds on \mh, \mA\ and \tanb\ for the 
\mh-max benchmark.
The full lines represent the actual observation and
the dashed lines the limits expected on the basis of `background only' 
Monte Carlo experiments.
Upper right: projection (\mh,\mA);
lower left: projection (\mh,\tanb); 
lower right: projection (\mA,\tanb).  
\label{mssm-maxmh}}
\end{center}
\end{figure}

The Minimal Supersymmetric extension of the Standard Model (MSSM)
is the most attractive extension of the SM. The LEP experiments have
searched for the reactions $\rm \ee~\ra~hA~\ra~\bb\bb$ and $\bb\tautau$.
No indication of a signal has been observed 
as shown in Fig.~\ref{mssm-mass}
for the example of $\mh\approx\mA$, and the
resulting confidence levels $CL_b$ and $CL_s$ are given.
Figure~\ref{mssm-maxmh} shows the so-called benchmark results 
in the MSSM for large mixing in the scalar top sector (\mh-max)
for $CL_b$ and $CL_s$, leading to mass limits of
88.3 and 88.4~\Gcs\ on the CP-even and CP-odd neutral Higgs bosons,
and to an exclusion of the range $0.7<\tanb<1.8$ at 95\% CL~\cite{moriond00}.

\clearpage
\section{A General MSSM Parameter Scan} 

Important reductions of the mass limits compared with benchmark results
were reported for LEP1 and LEP2 data~\cite{as}.
With increasing statistics the reduction was only 6 to 8~\Gcs\ including 
the 189 GeV data of one LEP experiment (DELPHI)~\cite{delphi189} and similar for OPAL~\cite{opal189}.
Figure~\ref{mssm-maxmh} shows new results from a MSSM parameter
scan including the combined LEP data up to 202~GeV
for $CL_b$ and $CL_s$, leading to mass limits of
86 and 87~\Gcs\ on the CP-even and CP-odd neutral Higgs bosons,
and an exclusion of the range $0.7<\tanb<1.8$ at 95\% CL.
These limits are almost identical to the benchmark limits 
(Fig.~\ref{mssm-maxmh}),
which is expected for the high luminosity of the combined data.

\begin{figure}[htp]
\vspace*{-0.2cm}
\begin{center}
\epsfig{figure=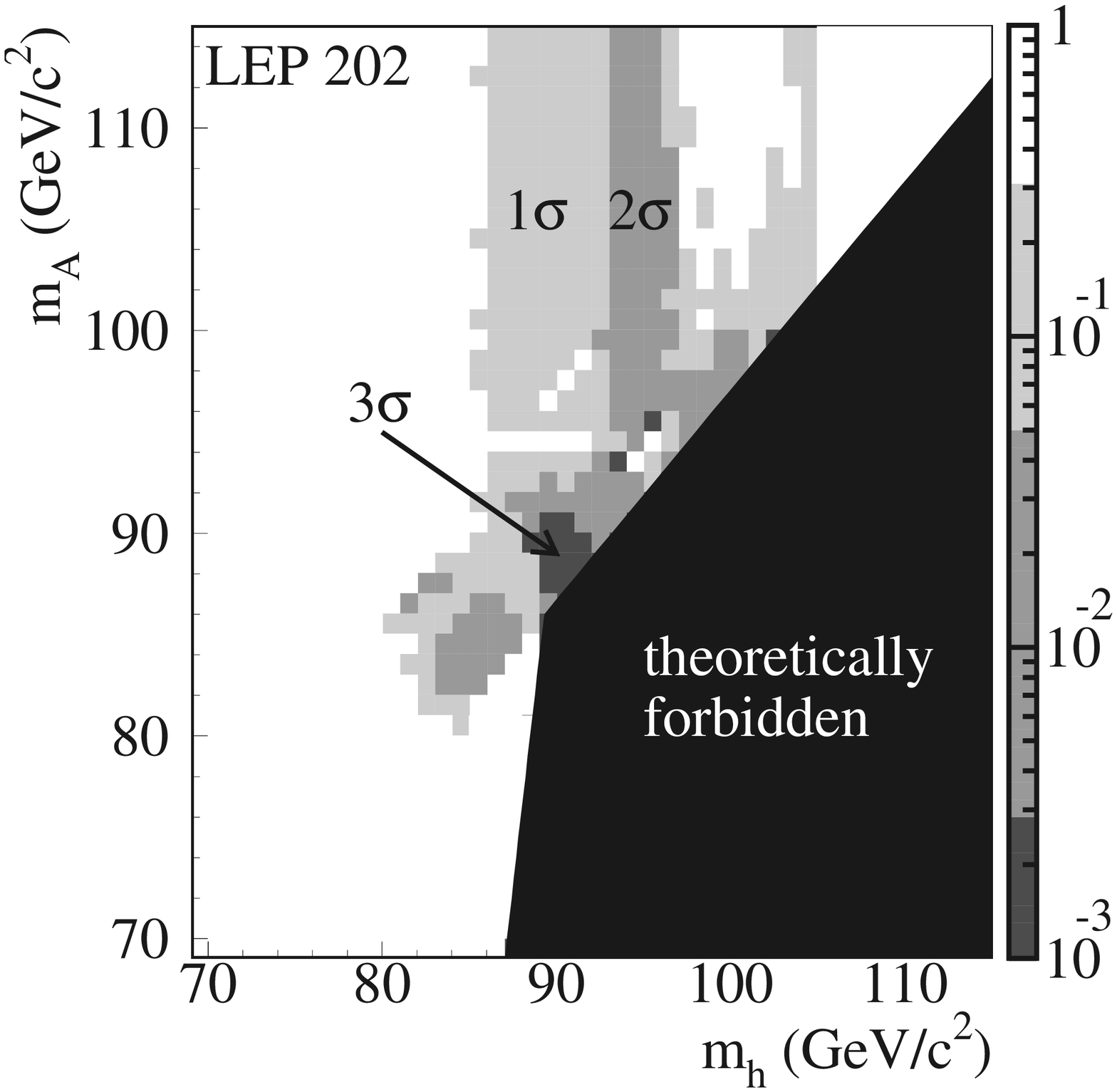,width=0.49\textwidth}
\epsfig{figure=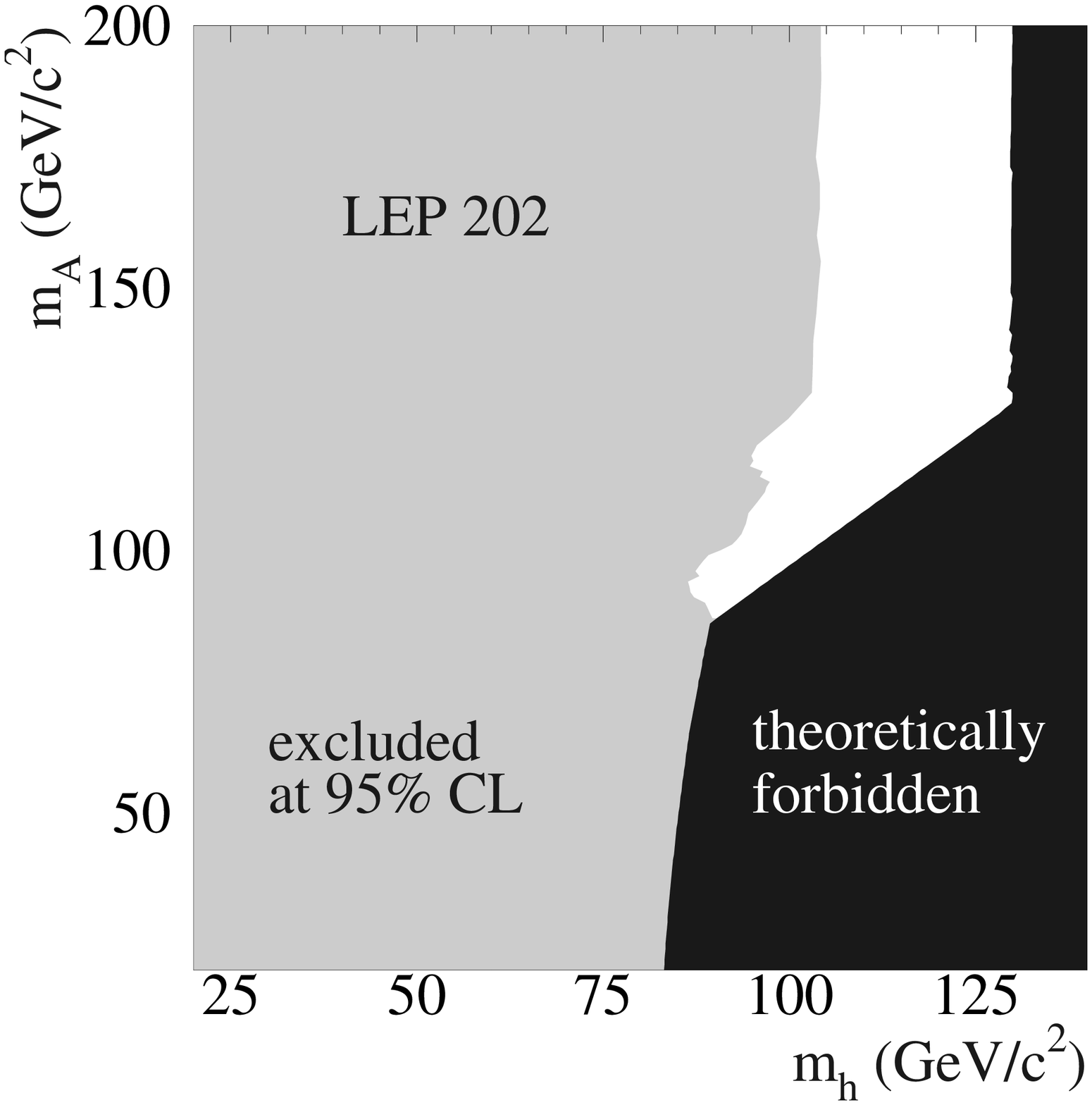,width=0.49\textwidth} 
\epsfig{figure=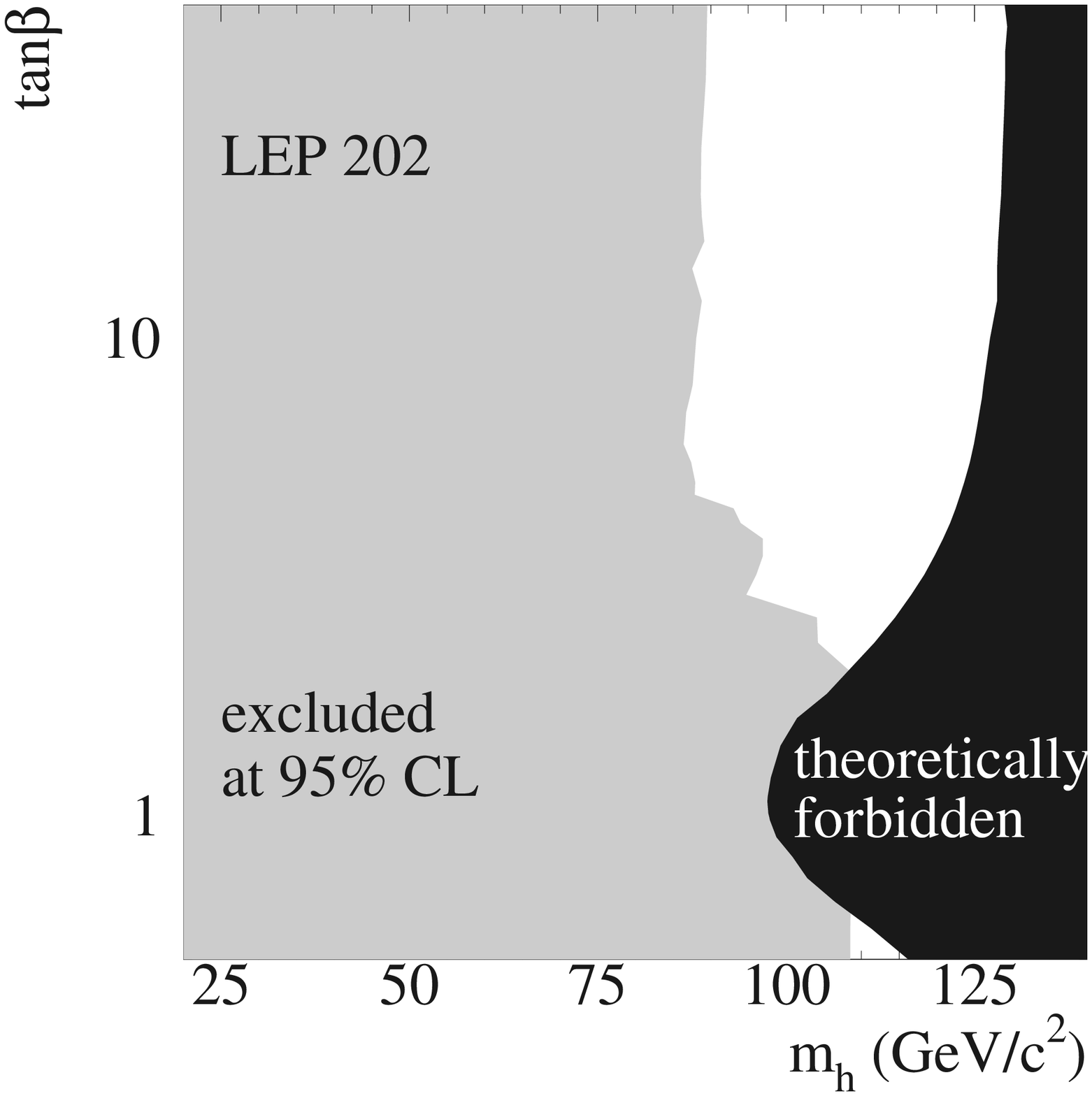,width=0.49\textwidth} 
\epsfig{figure=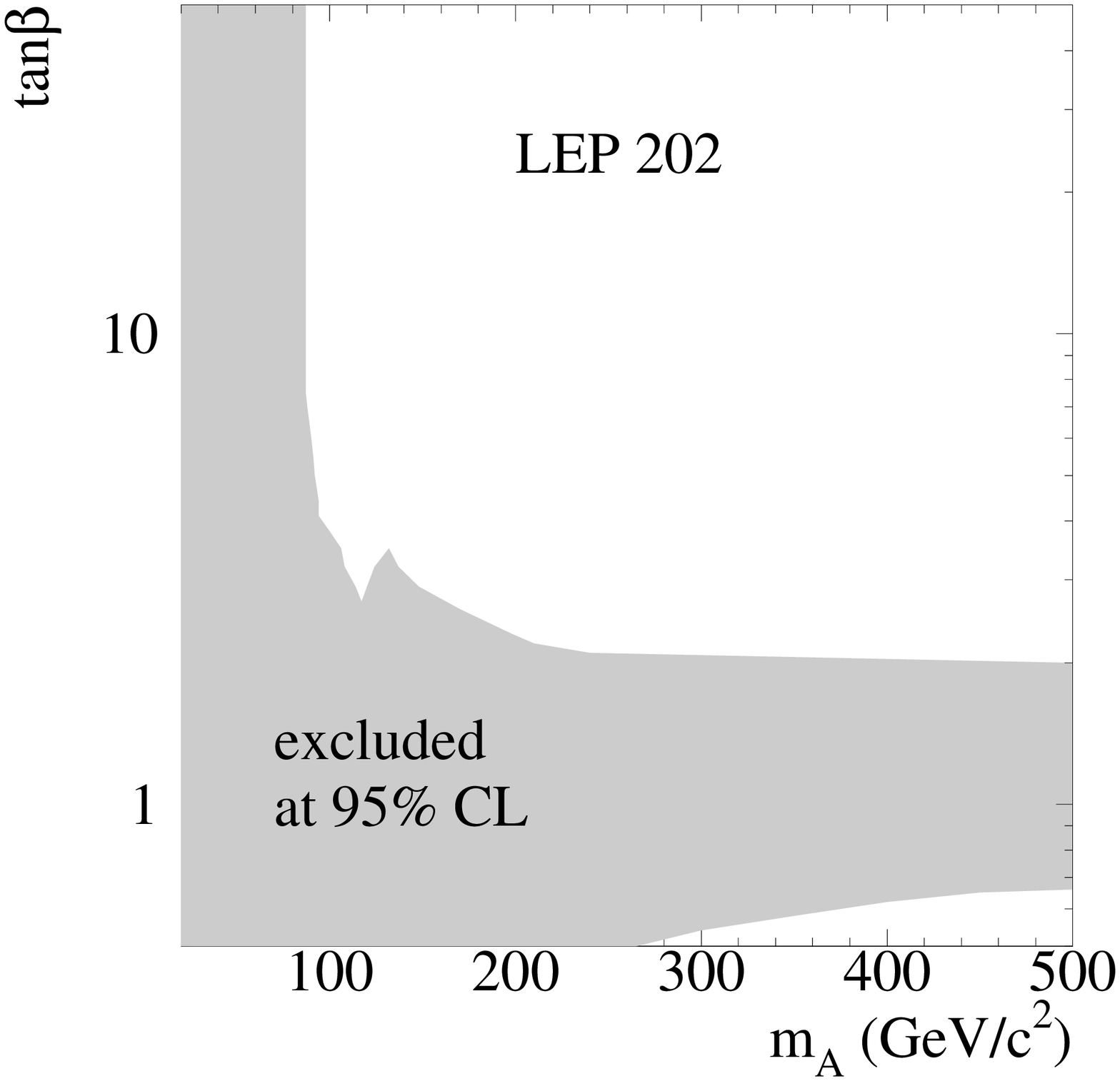,width=0.49\textwidth} 
\vspace*{-0.2cm}
\caption[]{\small 
MSSM parameter scan results in direct comparison with Fig.~4.
\label{mssm-scan}}
\end{center}
\vspace*{-0.2cm}
\end{figure}

\clearpage
\section{Charged Higgs Bosons}
\begin{figure}[htb]
\vspace*{-0.6cm}
\begin{center}
\begin{minipage}{0.48\textwidth}
\epsfig{figure=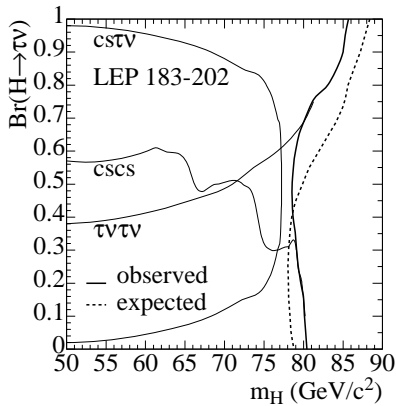,width=1.0\textwidth}
\end{minipage}
\begin{minipage}{0.48\textwidth}
\caption[]{\small 
The 95\% CL bounds on \mHpm\ as a function of the branching ratio
Br(\Hp~\ra~\tp$\nu$), combining the data collected by the four LEP 
experiments at energies from 183 to 202~GeV. 
The expected median exclusion limits are indicated by the dashed line and the
observed limits by the heavy full line. The light full lines show the 
observed limits channel by channel.
\label{charged-limit}}
\end{minipage}
\end{center}
\vspace*{-0.8cm}
\end{figure}

\vspace*{-0.1cm}
The search for charged Higgs bosons is performed in
the framework of the general extension of the SM with
two Higgs boson doublets. The combined results from the 
four LEP experiments for the reactions 
$\ee~\ra~\Hp\Hm~\ra~\csbar\cbars$, cs$\tau\nu$, and $\tp\nu\tm\nubar$
are presented in Fig.~\ref{charged-limit}, resulting in the 
limit of 78.6~\Gcs\ at 95\% CL, which is valid for any 
branching ratio Br(\Hp~\ra~\tp$\nu$).

\section{Conclusions}
\vspace*{-0.1cm}
The combination of the 1999 data from the four LEP experiments
resulted~in a large increase for the sensitivity of Higgs bosons;
however, no indication of a signal has been found. Stringent
mass limits are set for the SM Higgs boson, the neutral Higgs bosons
of the MSSM, and charged Higgs bosons. Owing to the large luminosity
of the combined data, a general scan of the MSSM parameters 
excludes neutral Higgs bosons below 86~GeV and the range
$0.7 < \tan\beta < 1.8$.

\section*{Acknowledgments}
\vspace*{-0.1cm}
I would like to thank the organizers of the conference
for their kind hospitality.

\end{document}